\documentclass{pasa}
\usepackage{graphicx}
\usepackage{times}
\pdfoutput=1   

\usepackage{aas_macros} 

\def\Mo{M$_\odot$}

\def\kms {\hbox{${\rm km\ s}^{-1}$}}
\def\ccm {$\hbox{{\rm cm}}^{-3}$}    
\def\scm  {$\hbox{{\rm cm}}^{-2}$}
\def \HI {HI}

\def \WpHz {W Hz$^{-1}$}
\def\lapp{\ifmmode\stackrel{<}{_{\sim}}\else$\stackrel{<}{_{\sim}}$\fi}
\def\gapp{\ifmmode\stackrel{>}{_{\sim}}\else$\stackrel{>}{_{\sim}}$\fi}
\title[Depletion of star-forming gas by AGN ]{The depletion of star-forming gas by AGN activity in radio sources}
\author[S. J. Curran]{S. J. Curran\thanks{Stephen.Curran@vuw.ac.nz}
\affil{School of Chemical and Physical Sciences, Victoria University of Wellington, PO Box 600, Wellington 6140, New Zealand}
}

\jid{PASA}
\doi{10.1017/pas.\the\year.xxx}
\jyear{\the\year}

\begin{document}

\begin{frontmatter} 
\maketitle

\begin{abstract}
Cold, neutral interstellar gas, the reservoir for star formation, is
  traced through the absorption of the 21-centimetre continuum
  radiation by neutral hydrogen (\HI).  Although detected in one
  hundred cases in the host galaxies of distant radio sources, only
  recently have column densities approaching the maximum value
  observed in Lyman-$\alpha$ absorption systems ($N_{\text{\HI}}\sim
  10^{22}$~\scm) been found. Here we explore the implications these
  have for the hypothesis that the detection rate of \HI\ absorption is
  dominated by photo-ionisation  from the active galactic nucleus
  (AGN).  We find, with the addition all of the current searches for
  \HI\ absorption at $z\geq0.1$, a strong correlation between
  the \HI\ absorption strength and the ionising photon rate, with the
  maximum value at which \HI\ is detected remaining close to the theoretical value
  in which all of the neutral gas would be ionised in a large spiral
  galaxy ($Q_\text{\HI} = 2.9\times10^{56}$ ionising photons s$^{-1}$).
  We also rule out other effects (excitation by the radio
  continuum and changing gas properties) as the dominant cause for
  the decrease in the detection rate with redshift. Furthermore, from the
  maximum theoretical column density we find that the five high
  column density systems have spin temperatures close to those of the
  Milky Way ($T_{\rm spin}\lapp300$~K), whereas, from our model of a
  gaseous galactic disk, the \HI\ detection at $Q_\text{\HI}
  =2.9\times10^{56}$~s$^{-1}$ yields $T_{\rm spin}\sim10\,000$~K,
  consistent with the gas being highly ionised. 
\end{abstract}

\end{frontmatter}

\section{INTRODUCTION}
\label{intro}

Since the first high redshift ($z\gapp3$) survey for cold neutral
(star-forming) gas, via the absorption of the 21-centimetre transition
of neutral hydrogen in the host galaxies  of distant radio sources, it
has been posited that the dearth of detections  
is due to the selection of ultra-violet (UV) luminous sources.
In these objects, the UV radiation from the AGN is sufficient to ionise
the gas to  below the detection limits of large radio telescopes \citep{cww+08}.
While a steady decrease with redshift, and hence UV
luminosity, may be expected, an abrupt cut-off in the detection of \HI\  above
$L_{\rm UV} \sim10^{23}$~\WpHz\ (ionising photon rates of 
$Q_\text{\HI}\gapp10^{56}$~s$^{-1}$) is apparent.
\citet{cw12} showed that such a critical luminosity would 
arise from an exponential gas distribution, with the observed  value being
close that required to ionise all of the gas in the Milky Way,
i.e. a large spiral. 

Since then, this observational result has been confirmed, not only
over specific ``homogeneous'' subsets of sources (compact, extended,
flat spectrum, etc., \citealt{cwt+12,caw+16,akk16,ak17,gdb+15,mmom21,mmko22}), but also
unbiased samples, limited only by flux
\citep{cwm+10,cwsb12,cwa+17,chj+17,chj+19,ace+12,gmmo14}.  The
complete ionisation of the gas within the host galaxies of these
objects would prevent star formation within them and is strongly
indicative of a selection effect, where the traditional need for a
reliable optical redshift, to which to tune the radio-band receiver, causes a
bias towards the most UV luminous sources \citep{cww+08}. This
suggests a population of undetected gas-rich galaxies in the distant
Universe, too faint to be detected via optical spectroscopy.

There is, however, still some debate over this interpretation:
\citet{cww+08} also noted that all of the sources above the critical
UV luminosity were type-1 objects (quasars), suggesting that the gas
could be undetected due to the obscuring circum-nuclear torus, invoked
by unified schemes of AGN \citep{ost78,am85,mg87}, not intercepting
our sight-line to the AGN. However, below the critical UV luminosity
the detection rate was similar to that in type-2 objects (galaxies),
suggesting that the bulk of the absorption occurs in the large-scale
galactic disc, which is randomly orientated to the pc-scale torus
\citep{cw10}.  Furthermore, rather than photo-ionisation from the AGN  being the dominant
cause of the decrease in detection rate with redshift,
\citet{ak17,ak18,aya+24} propose excitation of the hydrogen by
1.4~GHz photons \citep{pf56,fie59} or some other (unspecified)
evolutionary effect. While the former has been ruled out
(\citealt{cww+08,chj+19}, see also Sect.~\ref{asrl}), the latter effects would
have to apply across the whole sample, irrespective of source
classification in order to usurp the ionisation hypothesis.

Most recently, there have been five detections of \HI\ absorption
\citep{ckc20,mmom21,ssa+22,aya+24}, where the column density would exceed the
theoretical limit of $N_{\text{\HI}}\sim 10^{22}$~\scm\ \citep{sch01},
for moderate spin temperatures. In this paper we reassess the ionisation hypothesis in light of these
and the vastly increased sample of published \HI\ searches.

\section{Analysis}
\subsection{The data}

Adding the recent searches, comprising 441 objects
\citep{ckc20,mmom21,mmko22,mas+22,ssa+22,aya+24,dgc+24}, to those compiled in
\citet{chj+19} there are now 924 $z\geq0.1$ sources in the literature
which have been searched for associated \HI\ 21-cm absorption. These
are made up of 100 detections and 824 non-detections.

\subsection{Photometry and fitting}
\label{pandf}

\begin{figure}
  \includegraphics[scale=0.55]{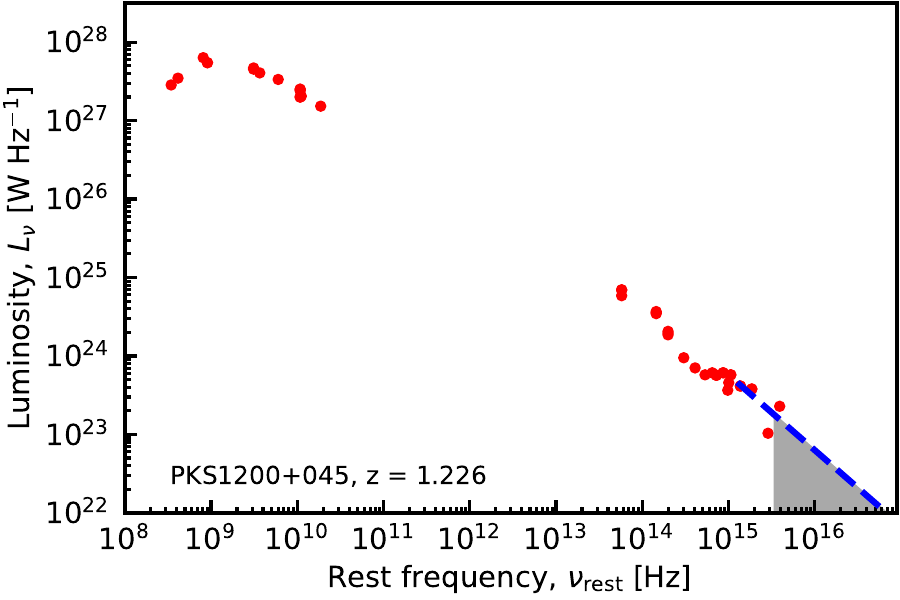}
  \vspace*{-0.1cm}
  \caption{Example of the rest-frame photometry. The dotted line shows the power-law fit to the UV data
    and the shaded
    region $\nu \geq 3.29\times10^{15}$~Hz over which the ionising photon rate is calculated. Here we show
    PKS\,1200+045, which with $Q_\text{\HI} = 2.9\times10^{56}$~s$^{-1}$ is the highest ionising photon
  rate at which \HI\  absorption has been detected (see Sect.~\ref{1200+045}).}
\label{ex_spec}
\end{figure}

To obtain the ionising photon rate for each of the 924 sources,
their photometry were scraped from the {\em NASA/IPAC
 Extragalactic Database} (NED), the {\em Wide-Field Infrared Survey
  Explorer} (WISE, \citealt{wem+10}) {\em Two Micron All Sky Survey}
(2MASS, \citealt{scs+06}) and the {\em Galaxy Evolution Explorer}
(GALEX, data release
GR6/7)\footnote{http://galex.stsci.edu/GR6/\#mission} databases.
After shifting the data back into the source's rest-frame, each flux
density measurement, $S_{\nu}$, was then converted to a specific
luminosity, via $L_{\nu}=4\pi \, D_{\rm L}^2 \,S_{\nu}/(z+1)$, where
$D_{\rm L}$ is the luminosity distance to the source (see Fig.~\ref{ex_spec}).\footnote{We use
$H_{0}=67.4$~km~s$^{-1}$~Mpc$^{-1}$ and $\Omega_{\rm m}=0.3125$
\citep{paa+20} throughout the paper.}
To obtain the ionising photon rate we use \citep{ost89},
\begin{equation}
Q_\text{\HI}\equiv\int^{\infty}_{\nu_{\rm ion}}\frac{L_{\nu}}{h\nu}\,d{\nu}, 
\end{equation}
where $\nu$ is the frequency (with $\nu_{\rm ion} = 3.29\times10^{15}$~Hz for \HI) and $h$ the Planck constant.
Fitting the rest-frame UV data with a power-law fit, $L_{\nu} \propto \nu^{\alpha}$, gives
\begin{equation}
\log_{10}L_{\nu} = \alpha\log_{10}\nu+ {\cal C},
\end{equation}
where $C$ is the log-space intercept and $\alpha$ the gradient (the spectral index).
Integrating this over $\nu_{\rm ion}$ to $\infty$ gives the ionising photon rate as
\begin{equation}
Q_\text{\HI} = \frac{-10^{\cal C}}{\alpha h}\nu_{\rm ion}^{\alpha},
\end{equation}
shown by the shaded region in Fig.~\ref{ex_spec}. In order to ensure a sufficient sample size, while
not contaminating the UV with optical-band data, we fit all photometry with $\log_{10}\nu \geq 15.1$
(shown by the dotted line in the figure), which left 180 sources with sufficient UV data.
Of these, 19 have been detected in \HI.

\subsection{\HI\ absorption strength}
\label{has}

The strength of the \HI\ 21-cm absorption is given by the profile's
velocity integrated optical depth ($\int\!\tau\,dv$), which is analogous to the
equivalent width in optical-band spectroscopy.  This is related to
the total neutral hydrogen column density via
\begin{equation}
N_{\text{\HI}}  =1.823\times10^{18}\,T_{\rm  spin}\int\!\tau\,dv,
\label{enew_full}
\end{equation}
where the spin temperature, $T_{\rm spin}$, quantifies the excitation from the lower hyperfine
level of the hydrogen atom \citep{pf56}.

We do not measure the intrinsic optical depth directly, but rather the observed optical depth,
$\tau_{\rm obs}$, which is the ratio of the line depth, $\Delta S$, to the observed
background flux, $S_{\rm obs}$. The two are related via
\begin{equation}
\tau \equiv-\ln\left(1-\frac{\tau_{\rm obs}}{f}\right) = -\ln\left(1-\frac{\Delta S}{fS_{\rm obs}}\right),
\label{tau_obs}
\end{equation}
where the covering factor, $f$, is the fraction of $S_{\rm obs}$ intercepted by the absorber.
In the optically thin regime, where $\tau_{\rm obs}\lapp0.3$, $\tau \approx \tau_{\rm obs}/f$,
so that  Equ. \ref{enew_full} can be approximated as
\begin{equation}
  N_{\text{\HI}}  \approx 1.823\times10^{18}\,\frac{T_{\rm  spin}}{f}\int\!\tau_{\rm obs}\,dv.
\label{enew}
\end{equation}
For the non-detections, the upper limit to the line strength is
obtained via $ \tau_{\rm obs} = {3\sigma_{\rm rms}}/S_{\rm obs}$, where
$\sigma_{\rm rms}$ is the rms noise level of the spectrum. In order to
place each of the limits on an equal footing each is re-sampled
to the same spectral resolution ($\Delta v = 20$~\kms), which is then
used as the FWHM to obtain the integrated optical depth limit per
channel (see \citealt{cur12}).

\begin{figure}
  \includegraphics[scale=0.58]{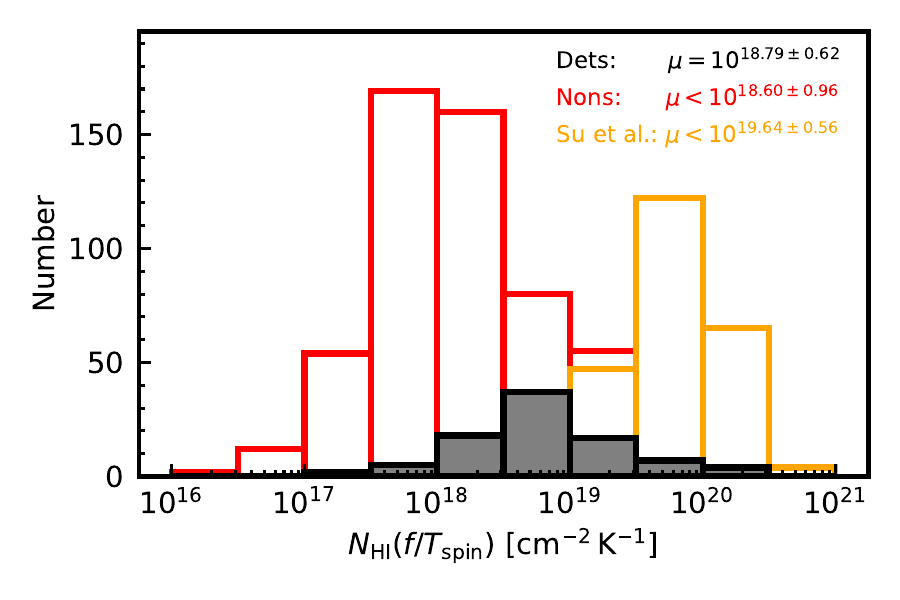}
  \vspace*{-0.8cm}
  \caption{The distributions of the normalised line strengths for the detections (filled histogram) and
    the upper limits (unfilled), which have been separated into the upper limits of \citet{ssa+22} and the
  rest of the sample. The legend shows the mean ($\pm1\sigma$) value of each distribution.}
\label{N-hist}
\end{figure}

Common practice is to convert the observed velocity integrated optical depth to a column density, by
assuming the spin temperature (and, presumably, $f=1$, e.g. \citealt{sgc+23}).
However, since we have no information on
this, or the covering factor\footnote{Where $N_{\rm  HI}$ is available, either from 21-cm emission at $z\lapp0.1$ or
Lyman-$\alpha$ absorption (at $z\gapp1.7$ with ground-based instruments), $T_{\rm  spin}/f$ can be measured,
although this varies greatly between objects: $40-300$~K within the Milky Way \citep{st04} and 
$10\lapp T_{\rm  spin}/f \lapp 10^4$~K at high redshift \citep{cur19},
as well across individual objects: In near-by galaxies this is $T_{\rm spin}/f \approx 2000$~K within
the stellar disk at galactocentric radii of $r\lapp10$~kpc before peaking 
at $T_{\rm spin}/f\approx 7000$~K at $r\approx15$~kpc \citep{cur20a}, where the OB stars are concentrated
\citep{mwc53}.}, we define the {\em normalised line strength}
\begin{equation}
1.823\times10^{18}\,\int\!\tau_{\rm obs}\,dv, \text{which gives }  N_{\text{\HI}}\left(\frac{f}{T_{\rm  spin}}\right).
\end{equation} 
Showing the distributions in Fig.~\ref{N-hist}, 
we see that while, in general, the non-detections have been searched sufficiently deeply to detect
\HI\ absorption, the sample of \citet{ssa+22} may not have been.\footnote{Most likely due to their
selection of very faint continuum sources ($S_{\rm obs}\gapp10$~mJy).} In order to reduce any
bias by including weaker limits, in the
rest of the analysis we only consider non-detections searched to $N_{\text{\HI}}\leq 10^{19}\,(T_{\rm spin}/f)$ \scm.

\section{Results and discussion}

\subsection{Factors affecting the detection of \HI}

\subsubsection{Source classification}
\label{cosm}

\begin{figure}
  \includegraphics[scale=0.58]{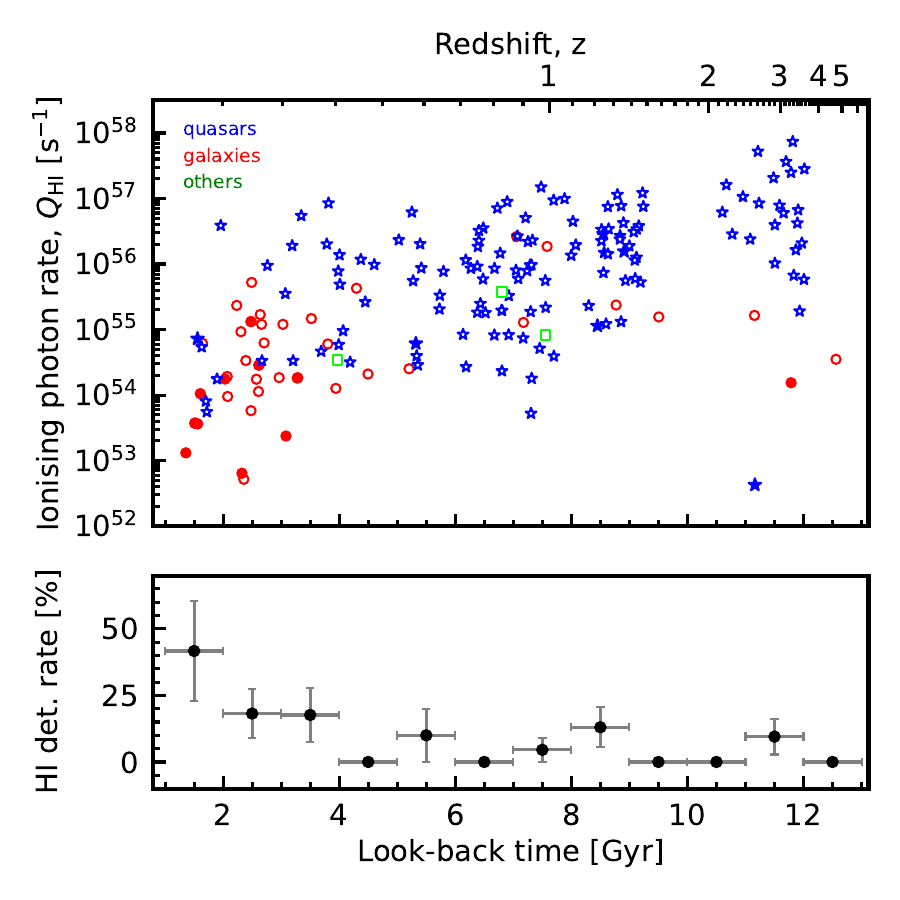}
  \vspace*{-0.8cm}
  \caption{The ionising photon rate versus the look-back time. The filled symbols show the \HI\ detections
    and the unfilled the non-detections, with the shapes designating the source classification:
    quasars -- {\em stars}, galaxies -- {\em circles}, other -- {\em squares}.
    The lower panel shows the \HI\ detection rate at
    various look-back times, where the error  bars on the ordinate show the Poisson standard errors and
    abscissa the range over which these apply.}
\label{Q-z}
\end{figure}

In Fig.~\ref{Q-z} we show the derived ionising photon rates versus the look-back time/redshift.
At low redshifts, we can see the high detection rate reported previously (e.g. \citealt{vpt+03,mmo+17}).
At best, we would expect a $\approx50$\% rate from the random
orientation of the absorbing medium, whether this be in the obscuring torus
or the large-scale galactic disk. However, it is clear that there is a
sharp decrease in the detection rate with redshift, which may be
caused by the preferential selection of type-1 objects (quasars),
where the AGN is not obscured by the torus.

\begin{figure}
  \includegraphics[scale=0.58]{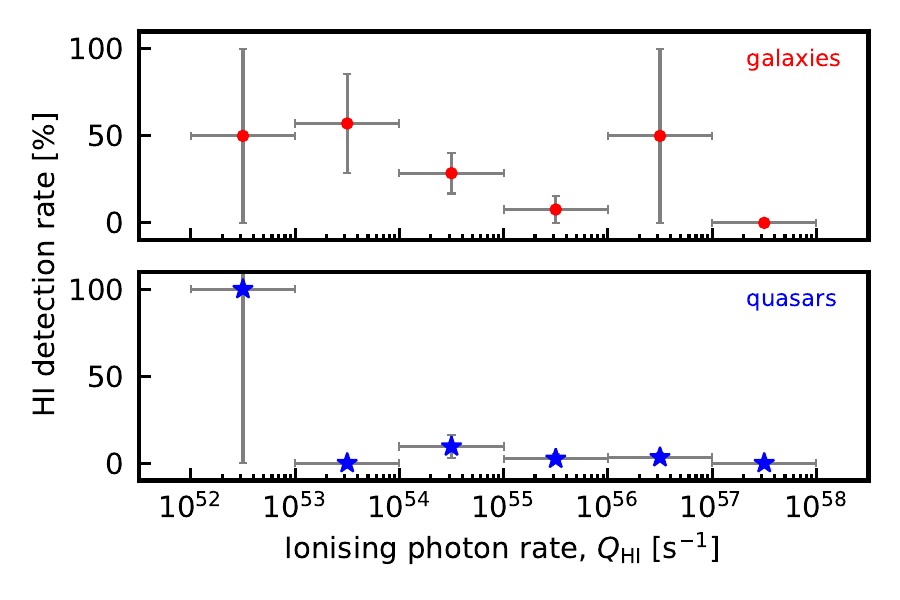}
 \vspace*{-0.8cm}
 \caption{The detection rate versus the ionising photon rate for the galaxies and the quasars. The error
   bars are as described in Fig.~\ref{Q-z}. For the galaxies the exact 50\% detection rates
   are due to a single detection and non-detection in the range and the 100\% detection rate
   for the quasars is due to only having a single object in the range. The \HI\ detected 
   galaxy in the $Q_\text{\HI} = 10^{56} - 10^{57}\text{ s}^{-1}$ bin is PKS\,1200+045 (see Sect.~\ref{1200+045}).}
\label{rate-Q}
\end{figure}

The galaxy and quasar detection rates are shown in Fig.~\ref{rate-Q}. Ignoring the 50 and 100\%
values, which comprise only one or two objects, we see that both detection
rates drop with increasing $Q_\text{\HI}$.  This is much steeper for
galaxies, although these start from much higher values.
Below the critical UV luminosity,
\citet{cww+08} found a $53\pm10$\% detection rate for galaxies and
$33\pm13$\% for quasars at $L_{\text UV} \lapp10^{23}$~\WpHz.  The current
numbers are smaller as we use the more stringent $\log_{10}\nu \geq
15.1$, cf.  $\log_{10}\nu \geq 14.8$, for the UV data, as well as the
ionising photon rate (by integrating the UV photometry), rather than
the monochromatic luminosity, which requires more complete UV
photometry. If we ignore the small number statistics\footnote{The
first three quasar bins have a total of just 4 quasars for
$Q_\text{\HI} < 10^{54}\text{ s}^{-1}$, 21 for $10^{54} < Q_\text{\HI}
< 10^{56}\text{ s}^{-1}$ and 40 for $10^{55} < Q_\text{\HI} <
10^{55}\text{ s}^{-1}$.}, this suggests that the orientation of the
torus may play a role, but given that quasars are nevertheless
detected in \HI\ absorption, this cannot be the whole story.  Note
also that \citet{mmko22} do not detect absorption in any of their 29
targets, considered to be galaxies and therefore expected to yield
several detections. However, the choice of targeting extended objects
may bias towards lowering covering factors, the effect of which is
suspected of reducing the observed optical depth 
in extended radio sources \citep{cag+13}.

Lastly, there is the simple explanation that quasars are generally more luminous than galaxies
(e.g. \citealt{ant93}) and, by scaling, have correspondingly higher ionisation rates.
\begin{figure}
  \includegraphics[scale=0.58]{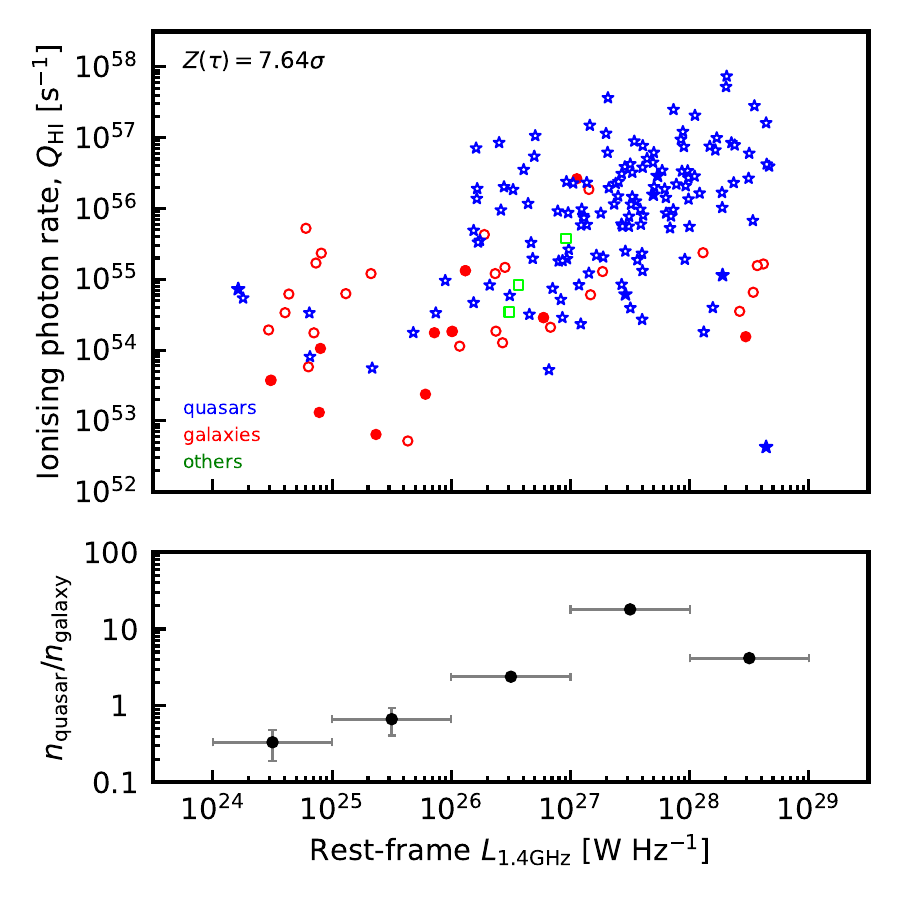}
 \vspace*{-0.7cm}
 \caption{The ionising photon rate versus the 21-cm continuum luminosity. The bottom panel shows the
   ratio of quasars to galaxies in each $L_{\text{1.4 GHz}}$ bin.}
\label{L-Q}
\end{figure}
This is apparent in Fig.~\ref{L-Q}, where the UV and radio luminosities are strongly correlated
[$p(\tau) = 2.17\times10^{-14}$]. That is, the Malmquist bias towards brighter objects at high redshift,
and fainter objects at low redshift, means that the brighter quasars will be more UV luminous resulting
in a lower \HI\ detection rate.

\subsubsection{Ionising photon rate}
\label{ipr}

\begin{figure}
  \includegraphics[scale=0.58]{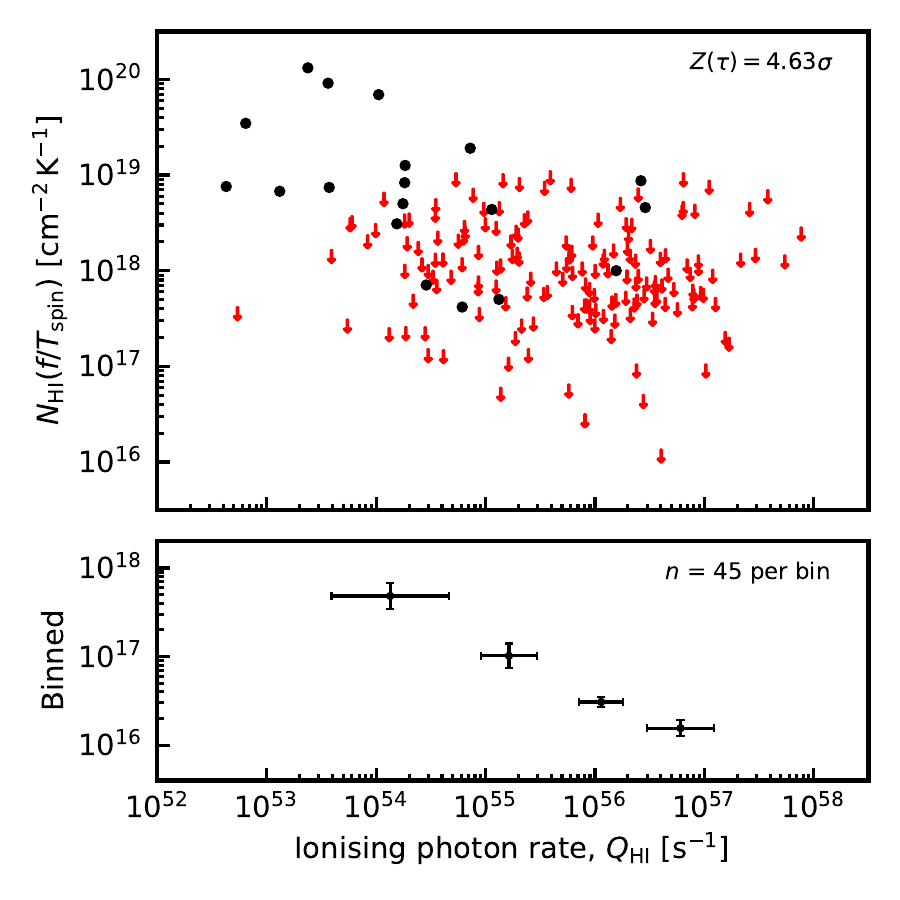}
  \vspace*{-0.7cm}
  \caption{The normalised absorption strength versus the ionising photon rate. The circles
    show the detections and the arrows the $3\sigma$ upper limits re-sampled to $\Delta v = 20$~\kms.
    The lower panel shows the data in equally sized bins with $\pm1\sigma$ error bars.}
\label{N-Q}
\end{figure}

In Fig.~\ref{N-Q}, we show the distribution of the normalised line
strengths versus the ionising photon rates for the sources which have
sufficient UV photometry.  For the 19 detections alone, a Kendall-tau
test gives a probability of $p(\tau) = 0.046$ for the
$N_{\text{\HI}}/({f}{T_{\rm spin}})-Q_{\text{\HI}}$ anti-correlation
occurring by chance. This is significant at $1.99\sigma$, assuming
Gaussian statistics.
If we include the upper limits, as censored data points, via
the {\em Astronomy SURVival Analysis} ({\sc asurv})
package \citep{ifn86}, the probability becomes
$p(\tau) = 3.66\times10^{-6}$ ($4.63\sigma$). 

Of the five absorbers with $N_{\text{\HI}} \gapp10^{20}\,(T_{\rm
  spin}/f)$ \scm, only one has sufficient UV photometry to obtain the
ionising photon rate (WISEA\,J145239.38+062738.2). With $Q_\text{\HI} = 2.4\times10^{53}$~s$^{-1}$,
this is well below the
$Q_\text{\HI}\sim10^{56}$~~s$^{-1}$ cut off. However, from the binning in
Fig.~\ref{N-Q}, the anti-correlation between absorption
strength and ionising photon rate is clear.\footnote{The limits are included via the
\citet{km58} estimator.} Lastly, below the maximum
detected value of $Q_\text{\HI} = 2.9\times10^{56}$~s$^{-1}$ there are
19 detections and 124 non-detections, giving a detection rate of
13.3\%. Above the maximum detected value, there are 40 non-detections
and, of course, 0 detections.  For $p=0.133$, the binomial probability
of obtaining 0 detections out of 40 at $Q_\text{\HI} > 2.9\times10^{56}$~s$^{-1}$ is
$3.32\times10^{-3}$ ($2.94\sigma$).

Bear in mind that there will be significant noise in these data, due
to different source sizes and morphologies (see Sect.~\ref{oeff}), and
the fact that a large fraction of the non-detections will simply not
be orientated favourably for us to detect absorption. This would
give, at best, a $\approx50$\% detection rate (Sect.~\ref{cosm}) and
if these could be removed, leaving only the ionisation as the factor
under consideration, we could see much more significant results.

\subsubsection{Radio luminosity}
\label{asrl}

As stated above, \citet{ak17,ak18} propose the  excitation of the hydrogen by 1.4~GHz photons
as a factor in the decrease in detection rate with redshift, although this was ruled out
by \citet{cww+08}. 
\begin{figure}
    \includegraphics[scale=0.58]{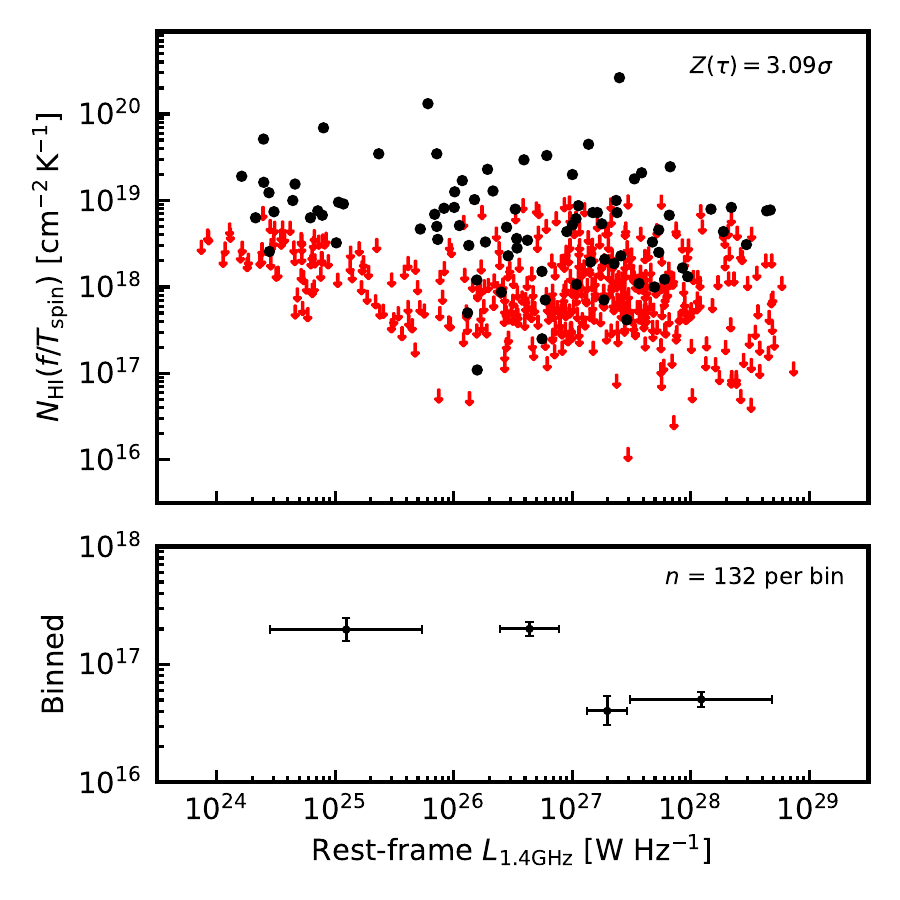}
  \vspace*{-0.7cm}
  \caption{As Fig.~\ref{N-Q}, but for the 21-cm continuum luminosity.} 
\label{L-N}
\end{figure}
Returning to this, in Fig.~\ref{L-N} we see that the \HI\ absorption strength also exhibits an
anti-correlation with the 21-cm continuum luminosity,
although with $p(\tau) = 0.0020$ this is considerably weaker than for the ionising photon rate. 
Furthermore, unlike for $Q_\text{\HI}$, it is seen that the detections and non-detections occupy a very
similar range of luminosities. Quantifying this, 
below the maximum
detected value of $L_{\text{1.4 GHz}} = 4.7\times10^{28}$~\WpHz\ there are
85 detections and 437 non-detections, giving a detection rate of
16.3\%. Above the maximum detected value, the 6 non-detections therefore give
a binomial probability of $0.344$ ($0.95\sigma$) of the distribution arising by chance.  
Thus, unlike the UV luminosity, there is no evidence of a critical radio luminosity, above
which \HI\ is not detected (as previously found by \citealt{cww+08,chj+19}).
We also note that a correlation between the line strength and radio luminosity would be expected
just from scaling with the ionising photon rate (see Fig.~\ref{L-Q}).

\subsubsection{Other effects}
\label{oeff}

\citet{ak17,ak18,aya+24} also propose other redshift evolutionary effects as the cause of the decrease
in detection rate with redshift. Regarding each of these:
\begin{itemize}
\item[--] {\em Source morphology}: It has long been known that the \HI\ absorption strength
  is anti-correlated with the size of the source \citep{pcv03}, which \citet{cag+13} suggested
  is a geometry effect introduced by the covering factor, and so we do expect
  higher detection rates in compact objects. However, given that \HI\ is detected over
  a range of source sizes, and neither compact nor non-compact objects are detected above the critical
  UV luminosity \citep{cw10}, the ionisation argument remains the more comprehensive. 

\item[--] {\em Gas properties}: Evolving gas properties could arise from either a changing column density 
  or evolving spin temperature (see Sects.~\ref{vh} \& \ref{1200+045}). Due to the weakness
  of \HI\ 21-cm emission, we do not usually have a measure of $N_{\text{\HI}}$ at $z\gapp0.1$, although,
  from the spectra of damped Lyman-$\alpha$ absorption systems (DLAs), there is no evidence
  of any evolution for intervening absorbers (\citealt{cur19} and references therein).
  Another possibility is an increase in the spin temperature of the gas (cf. the decrease in
  covering factor above). However, this would be expected to be a result of the high ionisation
  rates.
  
\item[--] {\em AGN luminosity}: Again, this would be a signature of the ionising photon rate,
  since we have ruled out the effect of the radio luminosity (Sect.~\ref{asrl}).
\end{itemize}
\citeauthor{ak18}  also propose an unspecified evolutionary effect. 
Due to the Malmquist bias, the ionising photon rate is strongly correlated with the redshift
(Fig.~\ref{Q-z}) and we can test this as above: The maximum redshift
at which \HI\ has been detected is $z=3.530$ \citep{ajj+20}. Below this redshift, there are 90
detections and 711 non-detections with $N_{\text{\HI}}\leq 10^{19}\,(T_{\rm spin}/f)$ \scm, giving
a detection rate of 11.2\%. Thus, the binomial probability
of obtaining 0 detections out of 12 at $z>3.530$  is $p=0.240$ ($1.17\sigma$). That is,
the detection of \HI\ appears to be much more dependent on the photo-ionisation than the redshift,
although both properties are intimately entwined.

\subsection{High column density systems}
\label{vh}

In Galactic high latitude clouds, above column densities of
$N_{\text{\HI}}\approx4\times10^{20}$~\scm\ \citep{rkh94,hbkw01}\footnote{This
limit is also apparent in the near-by Circinus galaxy \citep{ckb08}.}
the \HI\ begins to form H$_2$, with \citet{sch01} suggesting that this is the reason why high
redshift absorbers (DLAs) are never found with column
densities $N_{\text{\HI}}\gapp10^{22}$~\scm. Applying this to the current sample,
\citet{cw12} used a simple exponential model of the Galactic gas distribution, $n = n(0)e^{-r/R}$,
where $R$ is the scale-length of the decay. Extrapolating from 
$n = n_0 e^{-(r-R_{\odot})/R}$, where $n_0 = 0.9$~\ccm, $R_{\odot} = 8.5$~kpc and $R=3.15$~kpc
\citep{kk09} to $r=0$, gave $n(0)= 13.4$~\ccm.
The total column density between the continuum source and ourselves is therefore
\[
N_{\text{\HI}} = \int_0^{\infty} n dr = n(0)\int_0^{\infty}e^{-r/R} dr = n(0)R 
\]
$= 1.3\times10^{23}$~\scm, which is about an order of magnitude higher than that
expected. 

\begin{figure}
  \includegraphics[scale=0.58]{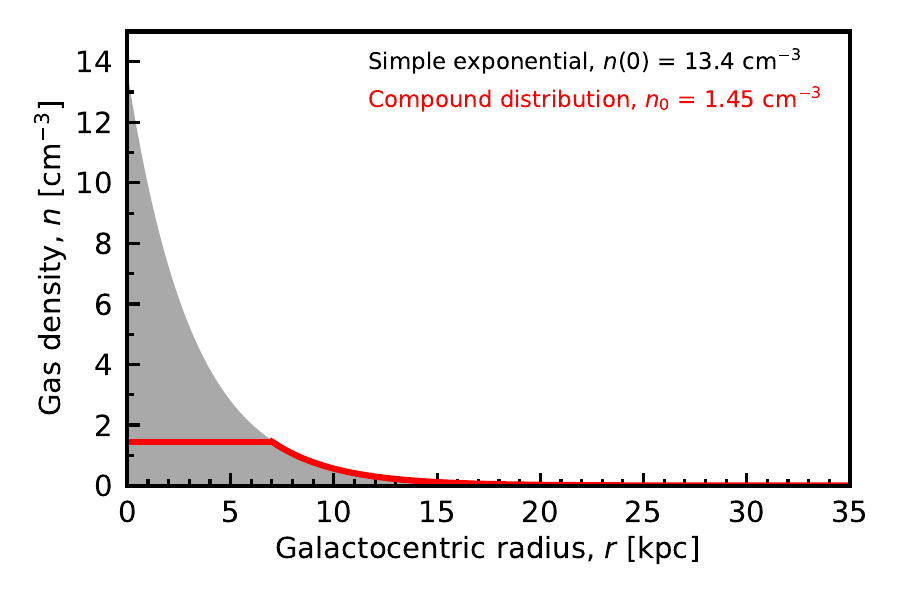}
  \vspace*{-0.7cm}
  \caption{The gas density versus the galactocentric radius for the
    simple exponential (grey) and compound (red) models of the Milky Way. $r_0$ is radius
  at which the break between the models occurs.}
\label{exp_model}
\end{figure}

We therefore proceed by using a compound model, where
a constant density component is added over $0\leq r \leq r_0$ 
to the exponential component (Fig.~\ref{exp_model}),
giving
\begin{equation}
N_{\text{\HI}} = \int_{0}^{r_0} n_0 dr +   \int_{r_0}^{\infty} n_0 e^{-(r-R_{\odot})/R} dr, 
\label{N1}
\end{equation}
where $n_0 = 0.9$~\ccm, $R= 3.15$~kpc, $R_{\odot} = 8.5$~kpc and $r_0= 7$~kpc \citep{kk09}.
To reduce the number of free parameters, we rewrite the formula of \citeauthor{kk09}
\[
n = n_0 e^{-(r-R_{\odot})/R} \text{ as } n =  n_0 e^{-(r-r_0)/R},
\]
where $n=0$ is now the gas density at $r_0$, e.g. $n_0 = 1.45$~\ccm\ at $r_0 = 7$~kpc,
cf.  $n_0 = 0.9$~\ccm\ at $R_{\odot} = 8.5$~kpc for the Milky Way (Fig.~\ref{exp_model}).
Making the substitution and integrating, Equ.~\ref{N1} becomes
\begin{equation}
N_{\text{\HI}} =  n_0\left(r_0 +  Re^{-r_0/R}\right),
\label{N2}
\end{equation}
giving  $N_{\text{\HI}}= 3.3\times10^{22}$~\scm, which is closer to the expected
limit.

Since this value is obtained through an inclined disk, we may expect it to be close to representing the
theoretical limit. For the five \HI\ absorbers with $N_{\text{\HI}} \gapp10^{20}\,(T_{\rm spin}/f)$ \scm\
the absorption is
optically thick and so the approximation in Equ.~\ref{enew} cannot be used unless $f=1$. In any case,
having $f<1$ would have the effect of decreasing the already low spin temperatures (Table~\ref{4-table}).
\begin{table}  
  \caption{The five $N_{\text{\HI}} \gapp10^{20}\,(T_{\rm spin}/f)$
    \scm\ absorbers \citep{ckc20,mmom21,ssa+22,aya+24}.
    $N_{\text{\HI}}/(f/T_{\rm spin})$ gives the normalised absorption
    strength, followed by the spin temperature for
    $N_{\text{\HI}}=3.3\times10^{22}$ \scm.}
  \begin{tabular}{@{}l c r r@{}}
    \hline\hline
    Source     &  $z$ & $N_{\text{\HI}}/(f/T_{\rm spin})$& $T_{\rm spin}$\\
\hline
WISEA\,J022928.93+004429.5 & 1.217 & $1.35\times10^{20}$  & $\lapp240$ \\
RCS\,01020400291 &  1.163 & $1.45\times10^{20}$ &  $\lapp230$\\
MRC\,0531--237 & 0.851 & $2.63 \times10^{20}$ &  $\lapp130$\\
SDSS\,J090331.57+010847.5 & 0.522 & $2.14 \times10^{20}$ &  $\lapp150$\\
WISEA\,J145239.38+062738.2 & 0.267 & $1.32 \times10^{20}$  & $\lapp250$\\
\hline\hline
\end{tabular}
\label{4-table}  
\end{table} 

Such spin temperatures are typical of the Milky Way ($T_{\rm spin}\lapp300$~K, \citealt{st04,dsg+09}),
but may be atypical in sources host to a powerful AGN.
As mentioned in Sect.~\ref{cosm}, the ionising photon rate is only available for one of these
(WISEA\,J145239.38+062738.2), which has $Q_\text{\HI} = 2.4\times10^{53}$~s$^{-1}$. This is
three orders of magnitude below the highest value where \HI\ has been detected. 
Furthermore, this, and the other three high column
density systems, are at redshifts $z\ll 3$, meaning that the optical-band observation which yielded
the redshift are not close to the rest-frame UV band. This was identified as introducing
a possible bias by \citet{cww+08}, where the selection of 
objects faint in the $B$-band at $z\gapp3$, which were sufficiently bright to yield an optical redshift,
selected only objects which were very UV luminous in the source rest-frame.

Of the three high redshift exceptions where \HI\ has been detected, two\footnote{J0414+0534 at $z = 2.636$,
\citep{mcm98} and 0902+34 at $z = 3.398$ 
\citep{ubc91}.} have relatively low photo-ionisation rates
($Q_\text{\HI} \lapp 10^{54}$~s$^{-1}$, see Fig.\ref{Q-z}). For these, the redshifts were obtained from
spectroscopy of the near-infrared band (\citealt{lco95,lla85}, respectively),
thus remaining clear of the rest-frame
$\lambda\leq1216$~\AA\ range, where the hydrogen becomes excited and subsequently ionised.
For the other $z\gapp3$ detection (8C\,0604+728 at $z = 3.530$, \citealt{ajj+20}),
the redshift was obtained by deep optical observations towards a previously identified
radio source \citep{jwp+06}.\footnote{\citet{ajj+20}
claim $Q_\text{\HI}\approx2-5\times10^{56}$~s$^{-1}$ for this source,
although our photometry search could only find a single value with $\log_{10}\nu \geq 15.1$.
in the rest-frame. Nevertheless, this ionising photon rate
remains in the ballpark of the critical value.}

\subsection{\HI\ 21-cm absorption at the highest ionisation rate}
\label{1200+045}

From our fitting, the  highest ionising photon
rate at which \HI\ absorption has been detected \citep{ak17} occurs at $Q_\text{\HI} = 2.9\times10^{56}$~s$^{-1}$, in PKS\,1200+045 at $z=1.226$ (Fig.~\ref{N-Q}). This is the same as the theoretical 
$Q_\text{\HI}$ required to ionise all of the gas in the Milky Way \citep{cw12}. However, this
was based on the simple exponential distribution, which we have shown to overestimate
the column density (Sect.~\ref{vh}).

To obtain a revised value of the critical ionising photon rate, we again
start with the ionisation and recombination of the gas in equilibrium \citep{ost89},
\begin{equation}
Q_{\text{\HI}}= 4\pi\int^{r_{\rm str}}_{0}\,n_{\rm p}\,n_{\rm e}\,\alpha_{A}\,r^2\, dr
\label{Q1}
\end{equation}
where $n_{\rm p}$ and $n_{\rm e}$ are the proton and electron
densities, respectively, and $\alpha_{A}$ the radiative recombination
rate coefficient of hydrogen.  We use here the canonical $T=10\,000$~K
for ionised gas, giving
$\alpha_{A}=4.19\times10^{-13}$~cm$^3$~s$^{-1}$
\citep{of06}.\footnote{http://amdpp.phys.strath.ac.uk/tamoc/DATA/RR/}
For a neutral plasma, $n_{\rm p}=n_{\rm e}=n$, and for complete ionisation
of the gas ($r_{\rm str}=\infty$), the compound model gives
\begin{equation}
Q_{\text{\HI}}= 4\pi \alpha_{A} n_0^2 \left(\left[\frac{r^3}{3}\right]_0^{r_0} + \int^{\infty}_{r_0} e^{-2(r-r_0)/R} r^2\, dr\right)
\end{equation}
\begin{equation}
~~~~~~= \pi \alpha_{A} n_0^2 \left(\frac{4r_0^3}{3} + R\left[2r_0^2 + 2Rr_0 + R^2\right]\right),
\label{Q3}
\end{equation}
which for $r_0 = 0$ becomes the simple exponential model at large
radii, $Q_{\text{\HI}}= \pi \alpha_{A} n_0^2R^3$ \citep{cw12}. An
important feature of this is that the radius of the
Str\"{o}mgren sphere becomes infinite for a finite photo-ionisation
rate, giving the abrupt cut-off in \HI\ detections seen in the
observations.

From Equ.~\ref{Q3}, the ionising photon rate to ionise all of the gas
in the Milky Way is revised to $Q_\text{\HI} =
7.6\times10^{55}$~s$^{-1}$, which is a factor of four lower than for
the simple exponential model.  Of course all of gas need not be
ionised to be rendered below the detection limits of current radio
telescopes, although the apparently abrupt cut-off in the detection of
\HI\ at $Q_\text{\HI} \gapp 10^{56}$~s$^{-1}$ has persisted in all of
the published searches since \citet{cww+08} [see Sect.~\ref{intro}].

\begin{figure}
  \includegraphics[scale=0.58]{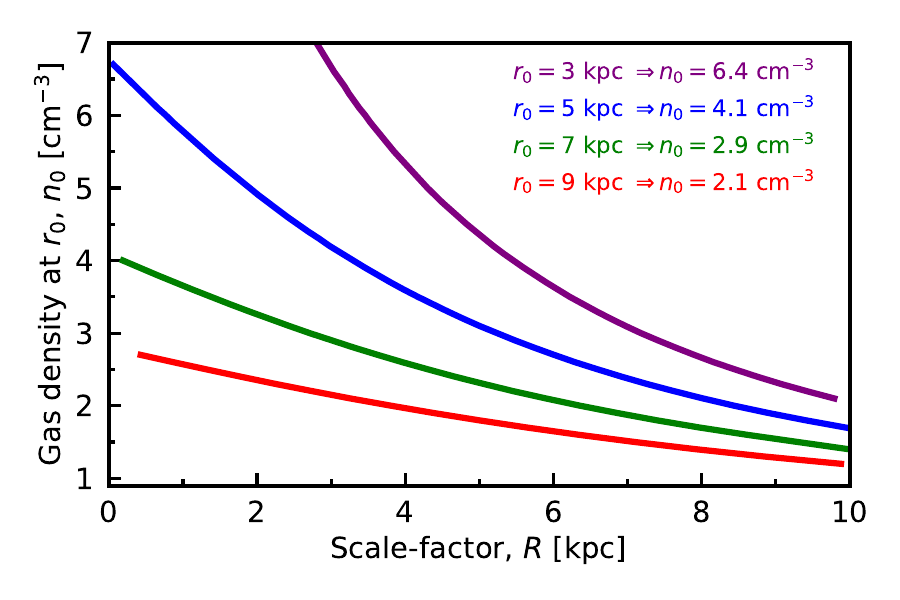}
  \vspace*{-0.7cm}
  \caption{The gas density at $r_0$ versus the scale-length required for all of the gas to be ionised
    by $Q_\text{\HI} = 2.9\times10^{56}$~s$^{-1}$. $r_0 = 7$~kpc for the Milky Way \citep{kk09} and the
  key shows the value of $n_0$ required for the Milky Way's $R=3.15$~kpc.}
\label{n0-R}
\end{figure}

In Fig.~\ref{n0-R} we show the ``tweaking'' required to the Galactic gas distribution to increase
the critical value to $Q_\text{\HI} = 2.9\times10^{56}$~s$^{-1}$.
For example, for the same values of $R$ and $r_0$ as the Milky Way, the central density
would have to be doubled to $n_0=2.9$~\ccm. Conversely, keeping $n_0=1.45$~\ccm\
gives the values of $r_0$ and $R$ listed in Table~\ref{2-table}.
\begin{table}  
  \caption{The required scale-length for various values of $r_0$ to yield complete ionisation
    for $Q_\text{\HI} = 2.9\times10^{56}$~s$^{-1}$ and $n_0=1.45$~\ccm. The column densities are
    calculated from Equ.~\ref{N2} and $T_{\rm spin}/f$ from the  measured
    $N_{\text{\HI}} = 4.6\times10^{18}\,(T_{\rm spin}/f)$~\scm\ \citep{ak17}.
    The gas masses are calculated from Equ.~\ref{M2} using the Galactic flare factor \citep{kk09}.}
\begin{tabular}{@{}r r r r r  r @{}}
 \hline\hline
\smallskip
$r_0$ [kpc] & $R$ [kpc] & $N_{\text{\HI}}$ [\scm] & $T_{\rm spin}/f$ [K] &  $M_{\rm gas}$ [\Mo] \\
\hline
$<7$ & $>10$ & $\lapp4\times10^{22}$ & $\lapp9000$  & $\gapp4\times10^{10}$\\
7.0 & 9.3 & $5.3\times10^{22}$ & 12\,000  & $3.5\times10^{10}$ \\
8.0 & 8.2 & $5.1\times10^{22}$  & 11\,000 & $3.3\times10^{10}$\\ 
9.0 & 7.1 & $5.1\times10^{22}$& 11\,000  & $3.1\times10^{10}$ \\
10.0 & 5.9 & $5.1\times10^{22}$  & 11\,000 & $2.8\times10^{10}$\\            
11.0 & 4.5 & $5.3\times10^{22}$  & 12\,000  & $2.4\times10^{10}$ \\              
12.0 & 3.0 & $5.6\times10^{22}$& 12\,000  & $1.9\times10^{10}$\\              
$\gapp13$ & $\lapp1$ & $\gapp6\times10^{22}$ & $\gapp13\,000$ & $\lapp 1\times10^{10}$\\
\hline
\end{tabular}
\label{2-table}  
\end{table}
From these parameters we estimate column densities which are
approximately equal to the maximum expected (Sect.~\ref{vh}) and use
these to estimate possible $T_{\rm spin}/f$ values, all of which are
very high.  While low covering factors ($f\ll1$) could contribute to
these, high spin temperatures would be expected from the strong UV
continuum \citep{fie59,be69}.\footnote{With an absorption strength of
$N_{\text{\HI}} = 7.8\times10^{18}\,(T_{\rm spin}/f)$ \scm, this model
yields $T_{\rm spin}/f\approx6000$~K for 8C\,0604+728 at $z = 3.530$
(see previous section).}

In the table we also show the total gas masses, obtained from $M_{\rm
  gas} = \int_0^{\infty}\rho dV$, where the volume of the disk gives
$dV = 2\pi tr dr$, with $t$ being its thickness. In the Galaxy the
thickness is related to the galactocentric radius via the flare
factor, $F = r/t\approx20$ \citep{kk09}, giving for the compound model
\begin{equation}
  M_{\rm gas}  = \frac{2\pi}{F} n_0 m_{\rm p}\left(\left[\frac{r^3}{3}\right]_0^{r_0} + \int_{r_0}^{\infty}e^{-(r-r_0)/R}r^2dr\right),
\label{M1}
\end{equation}  
which gives
\begin{equation}
M_{\rm gas} =  \frac{2\pi}{F} n_0 m_{\rm p}\left(\frac{r_0^3}{3} + R\left[r_0^2 + 2Rr_0 + 2R^2\right]\right).
\label{M2}
\end{equation}
The \HI\ masses derived from Equ.~\ref{M2} (Table~\ref{2-table}) are close to the maximum observed in 1000
low redshift galaxies ($M_{\rm gas} = 4\times10^{10}$~\Mo, \citealt{ksk+04}), indicating that
$Q_\text{\HI}\sim3\times10^{56}$~s$^{-1}$ is approaching the critical value above which all of the
gas in most galaxies will be ionised.

\section{Conclusions}

From the complete photometry of each of the 924 $z\geq0.1$ radio
sources searched for in \HI\ 21-cm absorption, we have collated the
ionising photon rates and radio luminosities, finding:
\begin{itemize}
\item The highest ionising photon rate at which \HI\ has been detected remains $Q_\text{\HI} \approx 3\times10^{56}$~s$^{-1}$,
  which is close to the value required to ionise all of the neutral gas in a large spiral galaxy, thus
  confirming that the dearth of \HI\ detections at high redshift is due to the bias towards sources which
  are most UV luminous in the rest-frame.
  
\item Both the ionising photon rate and radio luminosity are
  anti-correlated with the strength of the \HI\ absorption, although
  the $Q_{\text{\HI}}$ correlation is the strongest. Also, unlike the
  ionising photon rate, there is no critical radio luminosity above
  which \HI\ is not detected. That is, ionisation of the gas, rather
  than excitation to the upper hyper-fine level, appears to be the
  dominant mechanism for the dearth of \HI\ absorption at high
  redshift.

\item Any evolution in the source morphologies or gas properties
  cannot explain the decrease in detection rate with redshift as
  holistically as the ionisation hypothesis.
  
\item Detections rates are higher in galaxies than in
  quasars, which we attribute to the quasars generally being more
  luminous in the ultra-violet. It is possible that orientation
  effects play a role, although being a type-1 object does not
  necessarily exclude the detection of \HI\ absorption. This
  suggests that the absorption primarily occurs in the
  large-scale galactic disk, as opposed to the pc-scale obscuring
  torus.
 \end{itemize}  
From the total neutral hydrogen column density of the Milky Way \citep{kk09}, and that expected from theory, we find:
 \begin{itemize}  
  \item The strengths of the five recently detected \HI\ absorbers
    with $N_{\text{\HI}} \gapp10^{20}\,(T_{\rm spin}/f)$
    \scm\ \citep{ckc20,mmom21,ssa+22,aya+24}, imply spin temperatures
    of $T_{\rm spin}/f\lapp300$~K, which are typical of the Milky Way
    \citep{st04,dsg+09}. Sufficient UV photometry to obtain the
    ionising photon rate is only available for one of these, but with
    $Q_\text{\HI} = 2.4\times10^{53}$~s$^{-1}$ this is three orders of
    magnitude below the critical value above which we expect all of
    the gas to be ionised.

  \item Conversely, for the detection of \HI\ at the highest ionising photon rate
    ($Q_\text{\HI} =2.9\times10^{56}$~s$^{-1}$), we estimate $T_{\rm spin}/f\sim12\,000$~K which is consistent
    with a high ionisation fraction.

  \item At this ionising photon rate we calculate a gas mass of $M_{\rm gas} \approx3\times10^{10}$~\Mo, which is close to the maximum value observed in a survey of a 1000 low redshift galaxies \citep{ksk+04}.
    
    \end{itemize}  
The model is, of course, an idealisation, based upon
the gas distribution of the Milky Way and  taking no account of
shielding by dust\footnote{Which may be countered somewhat by the UV
photometry being uncorrected for dust, rendering the values as
relative rather than absolute.} or regions of denser gas
(e.g. molecular clouds). However, it is remarkable that it
comes close to yielding the maximum ionising photon rate at which
\HI\ has been detected for a gas distribution so similar to that of a
large spiral galaxy. Thus, both the extensive observational results and the
model suggest that ionisation by $\lambda\leq912$~\AA\ photons is the dominant
reason for the non-detection of cold, neutral gas within the host galaxies of high redshift
radio sources. 

\section*{Data availability}

Data available on request.

\section*{Acknowledgements}

I would like the thank the anonymous referee for their prompt and supportive feedback.
This research has made use of the NASA/IPAC Extragalactic Database
(NED) which is operated by the Jet Propulsion Laboratory, California
Institute of Technology, under contract with the National Aeronautics
and Space Administration and NASA's Astrophysics Data System
Bibliographic Service. This research has also made use of NASA's
Astrophysics Data System Bibliographic Service and {\sc asurv} Rev 1.2
\citep{lif92a}, which implements the methods presented in
\citet{ifn86}.


\begin{thebibliography}{}
\makeatletter
\relax
\def\mn@urlcharsother{\let\do\@makeother \do\$\do\&\do\#\do\^\do\_\do\%\do\~}
\def\mn@doi{\begingroup\mn@urlcharsother \@ifnextchar [ {\mn@doi@}
  {\mn@doi@[]}}
\def\mn@doi@[#1]#2{\def\@tempa{#1}\ifx\@tempa\@empty \href
  {http://dx.doi.org/#2} {doi:#2}\else \href {http://dx.doi.org/#2} {#1}\fi
  \endgroup}
\def\mn@eprint#1#2{\mn@eprint@#1:#2::\@nil}
\def\mn@eprint@arXiv#1{\href {http://arxiv.org/abs/#1} {{\tt arXiv:#1}}}
\def\mn@eprint@dblp#1{\href {http://dblp.uni-trier.de/rec/bibtex/#1.xml}
  {dblp:#1}}
\def\mn@eprint@#1:#2:#3:#4\@nil{\def\@tempa {#1}\def\@tempb {#2}\def\@tempc
  {#3}\ifx \@tempc \@empty \let \@tempc \@tempb \let \@tempb \@tempa \fi \ifx
  \@tempb \@empty \def\@tempb {arXiv}\fi \@ifundefined
  {mn@eprint@\@tempb}{\@tempb:\@tempc}{\expandafter \expandafter \csname
  mn@eprint@\@tempb\endcsname \expandafter{\@tempc}}}

\bibitem[\protect\citeauthoryear{{Aditya} \& {Kanekar}}{{Aditya} \&
  {Kanekar}}{2018a}]{ak17}
{Aditya} J.~N.~H.~S.,  {Kanekar} N.,  2018a, MNRAS, 473, 59

\bibitem[\protect\citeauthoryear{{Aditya} \& {Kanekar}}{{Aditya} \&
  {Kanekar}}{2018b}]{ak18}
{Aditya} J.~N.~H.~S.,  {Kanekar} N.,  2018b, MNRAS, 481, 1578

\bibitem[\protect\citeauthoryear{{Aditya}, {Kanekar}  \& {Kurapati}}{{Aditya}
  et~al.}{2016}]{akk16}
{Aditya} J.~N.~H.~S.,  {Kanekar} N.,   {Kurapati} S.,  2016, MNRAS, 455, 4000

\bibitem[\protect\citeauthoryear{{Aditya}, {Jorgenson}, {Joshi}, {Singh}, {An}
  \& {Chandola}}{{Aditya} et~al.}{2021}]{ajj+20}
{Aditya} J.~N.~H.~S.,  {Jorgenson} R.,  {Joshi} V.,  {Singh} V.,  {An} T.,
  {Chandola} Y.,  2021, MNRAS, 500, 998

\bibitem[\protect\citeauthoryear{{Aditya} et~al.,}{{Aditya}
  et~al.}{2024}]{aya+24}
{Aditya} J.~N.~H.~S.,  et~al., 2024, MNRAS, 527, 8511

\bibitem[\protect\citeauthoryear{{Allison} et~al.,}{{Allison}
  et~al.}{2012}]{ace+12}
{Allison} J.~R.,  et~al., 2012, MNRAS, 423, 2601

\bibitem[\protect\citeauthoryear{Antonucci}{Antonucci}{1993}]{ant93}
Antonucci R. R.~J.,  1993, ARA\&A, 31, 473

\bibitem[\protect\citeauthoryear{Antonucci \& Miller}{Antonucci \&
  Miller}{1985}]{am85}
Antonucci R. R.~J.,  Miller J.~S.,  1985, ApJ, 297, 621

\bibitem[\protect\citeauthoryear{{Bahcall} \& {Ekers}}{{Bahcall} \&
  {Ekers}}{1969}]{be69}
{Bahcall} J.~N.,  {Ekers} R.~D.,  1969, ApJ, 157, 1055

\bibitem[\protect\citeauthoryear{{Chowdhury}, {Kanekar}  \&
  {Chengalur}}{{Chowdhury} et~al.}{2020}]{ckc20}
{Chowdhury} A.,  {Kanekar} N.,   {Chengalur} J.~N.,  2020, ApJ, 900, L30

\bibitem[\protect\citeauthoryear{Curran}{Curran}{2012}]{cur12}
Curran S.~J.,  2012, ApJ, 748, L18

\bibitem[\protect\citeauthoryear{Curran}{Curran}{2019}]{cur19}
Curran S.~J.,  2019, MNRAS, 484, 3911

\bibitem[\protect\citeauthoryear{Curran}{Curran}{2020}]{cur20a}
Curran S.~J.,  2020, A\&A, 635, A166

\bibitem[\protect\citeauthoryear{Curran \& Whiting}{Curran \&
  Whiting}{2010}]{cw10}
Curran S.~J.,  Whiting M.~T.,  2010, ApJ, 712, 303

\bibitem[\protect\citeauthoryear{Curran \& Whiting}{Curran \&
  Whiting}{2012}]{cw12}
Curran S.~J.,  Whiting M.~T.,  2012, ApJ, 759, 117

\bibitem[\protect\citeauthoryear{Curran, Koribalski  \& Bains}{Curran
  et~al.}{2008a}]{ckb08}
Curran S.~J.,  Koribalski B.~S.,   Bains I.,  2008a, MNRAS, 389, 63

\bibitem[\protect\citeauthoryear{Curran, {Whiting}, {Wiklind}, {Webb}, {Murphy}
   \& {Purcell}}{Curran et~al.}{2008b}]{cww+08}
Curran S.~J.,  {Whiting} M.~T.,  {Wiklind} T.,  {Webb} J.~K.,  {Murphy} M.~T.,
   {Purcell} C.~R.,  2008b, MNRAS, 391, 765

\bibitem[\protect\citeauthoryear{Curran et~al.,}{Curran et~al.}{2011}]{cwm+10}
Curran S.~J.,  et~al., 2011, MNRAS, 413, 1165

\bibitem[\protect\citeauthoryear{Curran, {Whiting}, {Sadler}  \&
  {Bignell}}{Curran et~al.}{2013a}]{cwsb12}
Curran S.~J.,  {Whiting} M.~T.,  {Sadler} E.~M.,   {Bignell} C.,  2013a, MNRAS,
  428, 2053

\bibitem[\protect\citeauthoryear{Curran, {Whiting}, {Tanna}, {Sadler}, {Pracy}
  \& {Athreya}}{Curran et~al.}{2013b}]{cwt+12}
Curran S.~J.,  {Whiting} M.~T.,  {Tanna} A.,  {Sadler} E.~M.,  {Pracy} M.~B.,
  {Athreya} R.,  2013b, MNRAS, 429, 3402

\bibitem[\protect\citeauthoryear{{Curran}, {Allison}, {Glowacki}, {Whiting}  \&
  {Sadler}}{{Curran} et~al.}{2013c}]{cag+13}
{Curran} S.~J.,  {Allison} J.~R.,  {Glowacki} M.,  {Whiting} M.~T.,   {Sadler}
  E.~M.,  2013c, MNRAS, 431, 3408

\bibitem[\protect\citeauthoryear{Curran, {Allison}, {Whiting}, {Sadler},
  {Combes}, {Pracy}, {Bignell}  \& {Athreya}}{Curran et~al.}{2016}]{caw+16}
Curran S.~J.,  {Allison} J.~R.,  {Whiting} M.~T.,  {Sadler} E.~M.,  {Combes}
  F.,  {Pracy} M.~B.,  {Bignell} C.,   {Athreya} R.,  2016, MNRAS, 457, 3666

\bibitem[\protect\citeauthoryear{Curran, {Whiting}, {Allison}, {Tanna},
  {Sadler}  \& {Athreya}}{Curran et~al.}{2017a}]{cwa+17}
Curran S.~J.,  {Whiting} M.~T.,  {Allison} J.~R.,  {Tanna} A.,  {Sadler} E.~M.,
    {Athreya} R.,  2017a, MNRAS, 467, 4514

\bibitem[\protect\citeauthoryear{Curran, {Hunstead}, {Johnston}, {Whiting},
  {Sadler}, {Allison}  \& Bignell}{Curran et~al.}{2017b}]{chj+17}
Curran S.~J.,  {Hunstead} R.~W.,  {Johnston} H.~M.,  {Whiting} M.~T.,  {Sadler}
  E.~M.,  {Allison} J.~R.,   Bignell C.,  2017b, MNRAS, 470, 4600

\bibitem[\protect\citeauthoryear{Curran, {Hunstead}, {Johnston}, {Whiting},
  {Sadler}, {Allison}  \& {Athreya}}{Curran et~al.}{2019}]{chj+19}
Curran S.~J.,  {Hunstead} R.~W.,  {Johnston} H.~M.,  {Whiting} M.~T.,  {Sadler}
  E.~M.,  {Allison} J.~R.,   {Athreya} R.,  2019, MNRAS, 484, 1182

\bibitem[\protect\citeauthoryear{{Deka} et~al.,}{{Deka} et~al.}{2024}]{dgc+24}
{Deka} P.~P.,  et~al., 2024, A\&A

\bibitem[\protect\citeauthoryear{{Dickey}, {Strasser}, {Gaensler}, {Haverkorn},
  {Kavars}, {McClure-Griffiths}, {Stil}  \& {Taylor}}{{Dickey}
  et~al.}{2009}]{dsg+09}
{Dickey} J.~M.,  {Strasser} S.,  {Gaensler} B.~M.,  {Haverkorn} M.,  {Kavars}
  D.,  {McClure-Griffiths} N.~M.,  {Stil} J.,   {Taylor} A.~R.,  2009, ApJ,
  693, 1250

\bibitem[\protect\citeauthoryear{{Field}}{{Field}}{1959}]{fie59}
{Field} G.~B.,  1959, ApJ, 129, 536

\bibitem[\protect\citeauthoryear{{Ger{\'e}b}, {Maccagni}, {Morganti}  \&
  {Oosterloo}}{{Ger{\'e}b} et~al.}{2015}]{gmmo14}
{Ger{\'e}b} K.,  {Maccagni} F.~M.,  {Morganti} R.,   {Oosterloo} T.~A.,  2015,
  A\&A, 575, 44

\bibitem[\protect\citeauthoryear{{Grasha}, {Darling}, {Bolatto}  \&
  {Stocke}}{{Grasha} et~al.}{2019}]{gdb+15}
{Grasha} K.,  {Darling} J.~K.,  {Bolatto} A. D.~{Leroy} A.,   {Stocke} J.,
  2019, ApJS, 245, 3

\bibitem[\protect\citeauthoryear{{Heithausen}, {Br{\"u}ns}, {Kerp}  \&
  {Weiss}}{{Heithausen} et~al.}{2001}]{hbkw01}
{Heithausen} A.,  {Br{\"u}ns} C.,  {Kerp} J.,   {Weiss} A.,  2001, in
  {Pilbratt} G.~L.,  {Cernicharo} J.,  {Heras} A.~M.,  {Prusti} T.,   {Harris}
  R.,  eds,  Vol. 460, The Promise of the Herschel Space Observatory. p.~431

\bibitem[\protect\citeauthoryear{{Isobe}, {Feigelson}  \& {Nelson}}{{Isobe}
  et~al.}{1986}]{ifn86}
{Isobe} T.,  {Feigelson} E.,   {Nelson} P.,  1986, ApJ, 306, 490

\bibitem[\protect\citeauthoryear{Jorgenson, Wolfe, Prochaska, Lu, Howk, Cooke,
  Gawiser  \& Gelino}{Jorgenson et~al.}{2006}]{jwp+06}
Jorgenson R.~A.,  Wolfe A.~M.,  Prochaska J.~X.,  Lu L.,  Howk J.~C.,  Cooke
  J.,  Gawiser E.,   Gelino D.~M.,  2006, ApJ, 646, 730

\bibitem[\protect\citeauthoryear{{Kalberla} \& {Kerp}}{{Kalberla} \&
  {Kerp}}{2009}]{kk09}
{Kalberla} P.~M.~W.,  {Kerp} J.,  2009, Ann. Rev. Astr. Ap., 47, 27

\bibitem[\protect\citeauthoryear{Kaplan \& Meier}{Kaplan \& Meier}{1958}]{km58}
Kaplan E.~L.,  Meier P.,  1958, J. Amer. Statist. Assoc., 53, 457

\bibitem[\protect\citeauthoryear{{Koribalski} et~al.,}{{Koribalski}
  et~al.}{2004}]{ksk+04}
{Koribalski} B.~S.,  et~al., 2004, AJ, 128, 16

\bibitem[\protect\citeauthoryear{{Lavalley}, {Isobe}  \&
  {Feigelson}}{{Lavalley} et~al.}{1992}]{lif92a}
{Lavalley} M.~P.,  {Isobe} T.,   {Feigelson} E.~D.,  1992, in BAAS. pp 839--840

\bibitem[\protect\citeauthoryear{{Lawrence}, {Cohen}  \& {Oke}}{{Lawrence}
  et~al.}{1995}]{lco95}
{Lawrence} C.~R.,  {Cohen} J.~G.,   {Oke} J.~B.,  1995, AJ, 110, 2583

\bibitem[\protect\citeauthoryear{{Lilly}, {Longair}  \&
  {Allington-Smith}}{{Lilly} et~al.}{1985}]{lla85}
{Lilly} S.~J.,  {Longair} M.~S.,   {Allington-Smith} J.~R.,  1985, MNRAS, 215,
  37

\bibitem[\protect\citeauthoryear{{Maccagni}, {Morganti}, {Oosterloo},
  {Ger{\'e}b}  \& {Maddox}}{{Maccagni} et~al.}{2017}]{mmo+17}
{Maccagni} F.~M.,  {Morganti} R.,  {Oosterloo} T.~A.,  {Ger{\'e}b} K.,
  {Maddox} N.,  2017, A\&A, 604, A43

\bibitem[\protect\citeauthoryear{{Mahony} et~al.,}{{Mahony}
  et~al.}{2022}]{mas+22}
{Mahony} E.~K.,  et~al., 2022, MNRAS, 509, 1690

\bibitem[\protect\citeauthoryear{{Miller} \& {Goodrich}}{{Miller} \&
  {Goodrich}}{1987}]{mg87}
{Miller} J.~S.,  {Goodrich} B.~F.,  1987, BAAS, 19, 695

\bibitem[\protect\citeauthoryear{{Moore}, {Carilli}  \& {Menten}}{{Moore}
  et~al.}{1999}]{mcm98}
{Moore} C.~B.,  {Carilli} C.~L.,   {Menten} K.~M.,  1999, ApJ, 510, L87

\bibitem[\protect\citeauthoryear{{Morgan}, {Whitford}  \& {Code}}{{Morgan}
  et~al.}{1953}]{mwc53}
{Morgan} W.~W.,  {Whitford} A.~E.,   {Code} A.~D.,  1953, ApJ, 118, 318

\bibitem[\protect\citeauthoryear{{Murthy}, {Morganti}, {Oosterloo}  \&
  {Maccagni}}{{Murthy} et~al.}{2021}]{mmom21}
{Murthy} S.,  {Morganti} R.,  {Oosterloo} T.,   {Maccagni} F.~M.,  2021, A\&A,
  654, A94

\bibitem[\protect\citeauthoryear{{Murthy}, {Morganti}, {Kanekar}  \&
  {Oosterloo}}{{Murthy} et~al.}{2022}]{mmko22}
{Murthy} S.,  {Morganti} R.,  {Kanekar} N.,   {Oosterloo} T.,  2022, A\&A, 659,
  A185

\bibitem[\protect\citeauthoryear{{Osterbrock}}{{Osterbrock}}{1978}]{ost78}
{Osterbrock} D.~E.,  1978, Proc. Nat. Acad. Sci., 75, 540

\bibitem[\protect\citeauthoryear{Osterbrock}{Osterbrock}{1989}]{ost89}
Osterbrock D.~E.,  1989, Astrophysics of Gaseous Nebulae and Active Galactic
  Nuclei.
University Science Books, Mill Valley, California

\bibitem[\protect\citeauthoryear{{Osterbrock} \& {Ferland}}{{Osterbrock} \&
  {Ferland}}{2006}]{of06}
{Osterbrock} D.~E.,  {Ferland} G.~J.,  2006, {Astrophysics of Gaseous Nebulae
  and Active Galactic Nuclei}.
University Science Books, Sausalito, California

\bibitem[\protect\citeauthoryear{{Pihlstr{\" o}m}, {Conway}  \&
  {Vermeulen}}{{Pihlstr{\" o}m} et~al.}{2003}]{pcv03}
{Pihlstr{\" o}m} Y.~M.,  {Conway} J.~E.,   {Vermeulen} R.~C.,  2003, A\&A, 404,
  871

\bibitem[\protect\citeauthoryear{{Planck Collaboration} et~al.,}{{Planck
  Collaboration} et~al.}{2020}]{paa+20}
{Planck Collaboration} et~al., 2020, A\&A, 641, A6

\bibitem[\protect\citeauthoryear{{Purcell} \& {Field}}{{Purcell} \&
  {Field}}{1956}]{pf56}
{Purcell} E.~M.,  {Field} G.~B.,  1956, ApJ, 124, 542

\bibitem[\protect\citeauthoryear{{Reach}, {Koo}  \& {Heiles}}{{Reach}
  et~al.}{1994}]{rkh94}
{Reach} W.~T.,  {Koo} B.-C.,   {Heiles} C.,  1994, ApJ, 429, 672

\bibitem[\protect\citeauthoryear{Schaye}{Schaye}{2001}]{sch01}
Schaye J.,  2001, ApJ, 562, L95

\bibitem[\protect\citeauthoryear{{Skrutskie} et~al.,}{{Skrutskie}
  et~al.}{2006}]{scs+06}
{Skrutskie} M.~F.,  et~al., 2006, AJ, 131, 1163

\bibitem[\protect\citeauthoryear{{Strasser} \& {Taylor}}{{Strasser} \&
  {Taylor}}{2004}]{st04}
{Strasser} S.,  {Taylor} A.~R.,  2004, ApJ, 603, 560

\bibitem[\protect\citeauthoryear{{Su} et~al.,}{{Su} et~al.}{2022}]{ssa+22}
{Su} R.,  et~al., 2022, MNRAS, 516, 2947

\bibitem[\protect\citeauthoryear{{Su} et~al.,}{{Su} et~al.}{2023}]{sgc+23}
{Su} R.,  et~al., 2023, ApJ, 956, L28

\bibitem[\protect\citeauthoryear{Uson, Bagri  \& Cornwell}{Uson
  et~al.}{1991}]{ubc91}
Uson J.~M.,  Bagri D.~S.,   Cornwell T.~J.,  1991, PhRvL, 67, 3328

\bibitem[\protect\citeauthoryear{Vermeulen et~al.,}{Vermeulen
  et~al.}{2003}]{vpt+03}
Vermeulen R.~C.,  et~al., 2003, A\&A, 404, 861

\bibitem[\protect\citeauthoryear{{Wright} et~al.,}{{Wright}
  et~al.}{2010}]{wem+10}
{Wright} E.~L.,  et~al., 2010, AJ, 140, 1868

\makeatother
\end{thebibliography}

\label{lastpage}

\end{document}